# Inventions on menu and toolbar coordination
## A TRIZ based analysis

**Umakant Mishra**

Bangalore, India

http://umakantm.blogspot.in

**Contents**

## 1. Introduction

The menu is one of the most commonly used elements of a graphical user interface. The objective of a menu system is to provide various commands and functions to the user in an easy way so that the user can just select the desired operation from a given list instead of typing a complex command in the command prompt.

A toolbar also does a similar function as the menu but with certain differences. A menu has the advantage of holding a large number of items without needing any additional screen space. In contrast, each button on the toolbar permanently occupies some space on the screen. It's not possible to implement large number of functions through a toolbar, as they will occupy more and more valuable screen space. However, the toolbar has an advantage as it gives a single click access to any function unlike a menu system where the user has to navigate through sub-menus to ultimate discover the item he is looking for.

### 1.1 Similarities between menu and toolbar

- Both provide easy access to commands and functions through easy graphical interface.

- Both try to achieve advanced features like adaptability, user configurability, faster access, reduce confusion, enhance look and feel etc.

- Both of them are designed to be primarily operated though a pointing device like a mouse (although both of them may also support keyboard operations).

- The same options may be available both in menu and in toolbar.

### 1.2 Differences between menu and toolbar

- A toolbar permanently occupies screen space while the menu does not. More and more toolbars can occupy substantial amount of screen space. But increasing number of menu items does not occupy additional screen space.

- A toolbar provides a single click access to functions. But a menu bar requires several clicks to navigate through sub-menus to arrive at the desired option.

- Toolbars generally use more graphical icons than menu while drop down menus generally contain more textual contents than toolbar.

- Conventionally an advanced toolbar may contain only icons and no text, whereas an advanced menu may contain both textual description and icon.

## 2. Inventions on menu and toolbar coordination

As a menu and toolbar system shares many common objectives, it is often useful maintain some relationship to coordinate between both the elements of a GUI system. The relationships can be easy as both of them often share the same internal function. For example, the print option in a menu will (most likely) call the same function as the print button on the toolbar.

The following are some interesting inventions on menu and toolbar coordination selected from US patent database.

## 2.1 Method and system for adding buttons to a toolbar (5644739)

**Background**

The toolbars provide convenient alternatives to drop down menus. Some of the modern toolbars provide options to add and remove buttons from the toolbar, but through difficult and cumbersome methods. There is a need to provide an easy way of adding and removing buttons from a toolbar.

**Solution provided by the invention**

US Patent 5644739 (Invented by Elizabeth Moursund, Assigned to Microsoft Corporation, Jul 97) discloses a method of intuitively adding a button or other type of control to a toolbar. According to the invention dragging and dropping controls onto the toolbar can create new toolbar controls. The new control is bound to an operation of the object and can be executed by a mouse event.

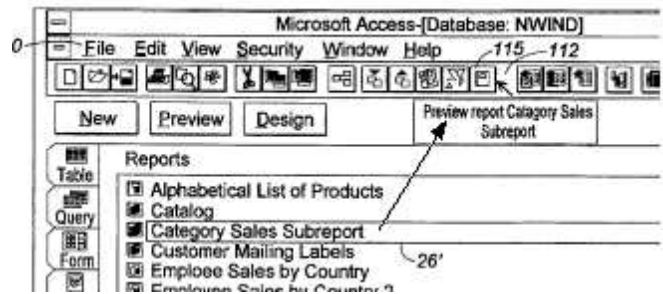

**TRIZ based analysis**

The invention allows dragging and dropping items on the toolbar to create buttons for the links. This method facilitates easy adding and removing buttons to the toolbar (Principle-15: Dynamize).

## 2.2 Representative mapping between toolbars and menu bar pulldowns (6177941)

Both pulldown menus and toolbars offer simple way of executing special commands. A pulldown menu can offer more user friendliness by indicating special keystrokes that can be used to select the pullcown commands even when the menu is not displayed. However, while scanning the menu bar pulldowns, it is not apparent whether there are corresponding toolbar items or not. And if the corresponding toolbar items do exist then there is no indication of where on the toolbar the items exist.

A similar problem is found in the reverse context. While selecting a toolbar icon, it is never apparent to a user whether there exist corresponding menu items or not. There is a need to devise a feedback mechanism to the users regarding the corresponding menu or toolbar icons while they are working on the toolbar or pulldown menus respectively.

**Solution provided by the invention**

Patent 6177941 (invented by Haynes, et al., assigned by IBM, issued Jan 2001) discloses a method of linking between a menubar pulldown and a toolbar icon. The method provides the visual correspondence when a graphical pointer is placed over an item on the pulldown menu or when a graphical pointer is placed over an item on the toolbar. The visual correspondence may be provided by highlighting the item on the toolbar, by providing a ghost image of the item on the toolbar, or by providing a ghost image of the pulldown menu.

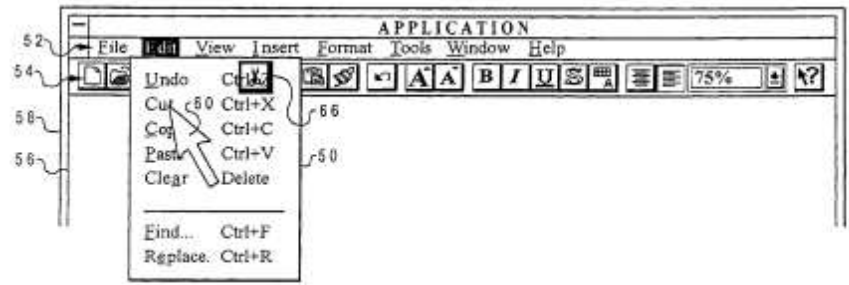

This method extends the functionality of the GUI by dynamically linking the menu bar and the toolbar without increasing visual complexity.

**TRIZ based analysis**

The invention improves the features of GUI by linking and indicating the corresponding menu bar and toolbar icons (Principle-38: Enrich).

According to the invention, moving pointer to a menu displays the corresponding toolbar icon and vice versa. (Principle-15: Dynamize).

The method provides a visual indication by highlighting the item on the toolbar, by providing a ghost image of the item on the toolbar or similar (Principle-32: Color change).

**2.3 Method for displaying controls in a system using a graphical user interface (6384849)**

**Background problem**

The GUI of an application program provides controls and functions through menus and toolbars. The user can choose the functions either through the menu or from the toolbar. Although toolbar and dropdown menus both provide ways to display controls, they look and feel very different. While toolbars have rich interactive controls, the drop down menu contains simple text strings. Both of them are treated differently and may contain different functions.

There is a need of an improved command bar that allows all controls to be included in either menu type containers or toolbar type containers.

**Solution provided by the invention**

Marcos et al. invented a method of displaying controls through a command bar (Patent 6384849, Assigned to Microsoft, May 02). According to the invention, the items may be displayed as both menu-like containers and toolbar like containers. The details of the control items are stored in a database. The data items include the Identification and description of the command bar along with their display state, i.e., whether menu-like or toolbar-like. The items are enabled for menu and/or for toolbar depending on their frequency of use.

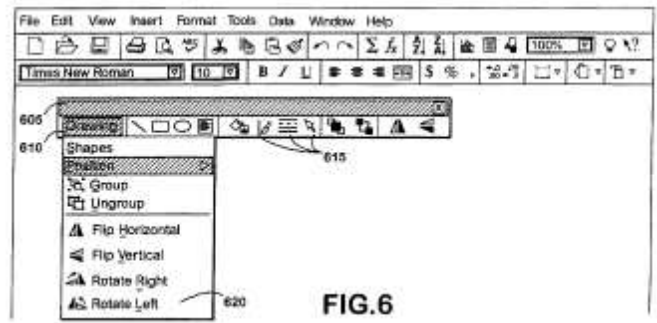

This invention of command bars integrate the features of both menubars and toolbars and provides methods for customizing them.

**TRIZ based analysis**

The command bar contains the functions and features of both menubar and toolbar (Principle-40: Composite).

The invention allows adding and removing menu popup from the command bar (Principle-15: Dynamize).

**2.4 Easy method of dragging pull-down menu items onto a toolbar (6621532)**

**Background problem**

Accessing an option through navigating a menu tree is time taking. This is worse in case of a sub-menu item which needs to activate and go through several level of menu by controlling the mouse pointer. Selecting an item on a toolbar is faster as it does not require activation of any menu. On the other hand a toolbar permanently occupies some real estate on the GUI.

**Solution provided by the invention**

US Patent 6621532 (invented by Mandt, assigned to IBM, Sep 03) provides a method of dragging pull down menu items onto a toolbar. According to the invention when the user drags a menu item and drops on the toolbar, the menu is automatically converted to a toolbar button. This facilitates the user to easily access the option during later use.

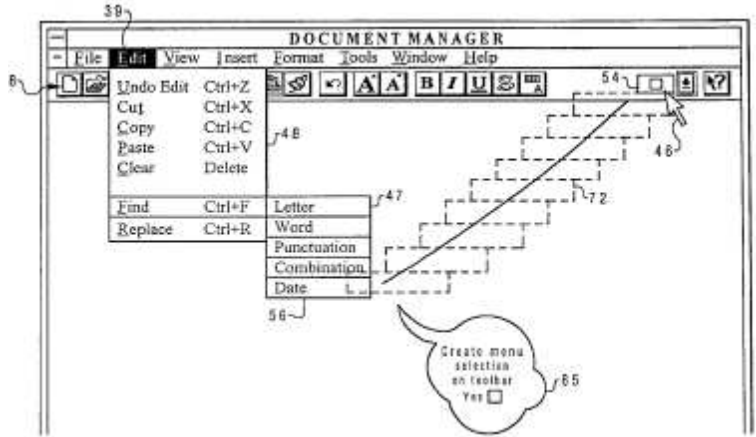

**TRIZ based analysis**

The menu mechanism consumes less screen space but navigation through a menu is difficult, as the user has to click several times. If menu items are displayed permanently like toolbar items, they will occupy more screen space (Contradiction).

The invention provides option to drag more frequently used option from a sub-menu to the toolbar. This method makes the item accessible by a single click (Principle-35: Parameter change).

## 3. Other related inventions

### 3.1 Method and apparatus for selecting button functions and retaining selected options on a display

Patent 5243697 (invented by Hoeber et al., assigned by Sun Microsystems, issued Sep 1993) discloses a method of using a pushpin to retain the frequently used menus on the display screen. According to the invention a frequently used menu can be kept open on the screen by using a pushpin option. This method has the advantages of menu and that of a toolbar as well. When used a pushpin the menu is converted to a permanent window and provides the benefit of a toolbar.

### 3.2 Procedural toolbar user interface

Patent 6456304 (Invented by Angiulo et al, Assigned to Microsoft, Sep 2002) provides a user interface toolbar that contains a plurality of dropdown menus. The options are all presented in menus, but the positioning is there on the toolbar. However, the presentation of menu options is context sensitive and based on the previous selections unlike a conventional menu.

## 4. Summary

Both toolbar and dropdown menu are used popularly in a graphical user interface with a similar objective of providing easy access to the internal functions. Often the same functions are provided through both menu and toolbar.

Both toolbar and dropdown menu have their own advantages and disadvantages. A menu can provide more options occupying less real estate, while toolbar can provide a single click access without navigating through trees and branches.

It is desirable to have coordination between dropdown menu and toolbar to avail the benefits of both. The above-illustrated inventions try to achieve this objective in different ways. Some interesting observations are:

- Transferring a menu item to toolbar and vice versa.
- Facilitating drag and drop operations between toolbar and menu.
- Presenting drop down menus in the form of a toolbar and vice versa.
- Displaying corresponding icons in a toolbar while the user navigates through menu and vice versa.
- Placing the menu window permanently on a screen like a toolbar.

Proper coordination and interaction between toolbar and dropdown menu can make the GUI more effective. We can expect to find more interesting inventions in this context in future.